\documentclass[nocompress]{spie}  

\usepackage{framed}
\usepackage{amsmath,amsfonts,amssymb}
\usepackage{graphicx}
\usepackage{hyperref}

\title{Projected power iteration for network alignment}

\author[a]{Efe Onaran}
\author[b]{Soledad Villar}
\affil[a]{Tandon School of Engineering, New York University}
\affil[b]{Courant Institute of Mathematical Sciences and Center for Data Science, New York University}

\authorinfo{Efe Onaran: eonaran@nyu.edu
\\  Soledad Villar: soledad.villar@nyu.edu}

\begin{document} 
\maketitle

\begin{abstract}
The network alignment problem asks for the best correspondence between two given graphs, so that the largest possible number of edges are matched. This problem appears in many scientific problems (like the study of protein-protein interactions) and it is very closely related to the quadratic assignment problem which has graph isomorphism, traveling salesman and minimum bisection problems as particular cases. The graph matching problem is NP-hard in general. However, under some restrictive models for the graphs, algorithms can approximate the alignment efficiently. In that spirit the recent work by Feizi and collaborators introduce EigenAlign, a fast spectral method with convergence guarantees for Erd\H{o}s-Reny\'i graphs. In this work we propose the algorithm Projected Power Alignment, which is a projected power iteration version of EigenAlign. We numerically show it improves the recovery rates of EigenAlign and we describe the theory that may be used to provide performance guarantees for Projected Power Alignment.  
\end{abstract}


\section{Introduction}
\label{sec:intro}  

Data science aims to extract information from data, usually by first proposing a model for the sought information, and then solving an optimization problem of the form 
\begin{equation}
\operatorname{minimize} f_D(x) \quad \text{ subject to } x\in \mathcal S. \label{eq.generic}
\end{equation}
In \eqref{eq.generic}, $D$ is the data for the problem, that can be thought to belong to a large set $\mathcal D$ of all possible problems; and $\mathcal S$ is the set of all feasible solutions for \eqref{eq.generic}.

For instance, the \textbf{network alignment} or \textbf{graph matching} problem consists in, given two graphs $G_1$ and $G_2$ with $n$ vertices, to find the best permutation to align $G_1$ with $G_2$. If $G_1$ and $G_2$ are isomorphic, then a solution is an isomorphism between them. If they are not, the solution is the permutation that matches the largest amount of edges possible between $G_1$ and $G_2$.  The network alignment problem has an optimization formulation of the form~\eqref{eq.generic} that we describe in Section \ref{sec:qap}, where $\mathcal S$ is the (discrete) set of $n\times n$ permutation matrices. Network alignment is a relevant problem for many scientific applications, including computational biology \cite{yang2007path}, computer vision \cite{rocha1994shape} and linguistics \cite{ehrig1992introduction}. In computational biology, network alignment is a useful tool for studying protein-protein interactions \cite{Kelley30092003}. Network alignment is closely related with the more general quadratic assignment problem that has the traveling salesman problem and the minimum bisection as particular cases. We give a broad description of network alignment and quadratic assignment in Section~\ref{sec:qap}, but for now we still consider the setting of the generic optimization problem~\eqref{eq.generic}.

The optimization problem \eqref{eq.generic} arising from data science is quite often NP-hard and therefore there is no hope to find an efficient algorithm that would find the exact solution for every $D$. However, being NP-hard is a statement about the hardest instance of the problem, and quite often, efficient algorithms can be used to solve or find an approximate solution for~\eqref{eq.generic} for a large subset of $\mathcal{D}$. Examples of such phenomenon include but are not restricted to convex relaxations. The approach may have started with the seminal work of Goemans and Williams \cite{goemans1995improved} and became very popular after compressed sensing.

Efficient algorithms may give satisfactory solutions to the problem \eqref{eq.generic} for $D\in \mathcal D'$ where $\mathcal D'$ is a proper subset of all possible data instances $\mathcal D$. Therefore it is natural (and a common practice within the applied mathematics, theoretical computer science and statistics communities) to study such algorithms under certain generative models. In those cases, the data is assumed to be generated by a specific probabilistic model which many times can be interpreted as 
\begin{equation}
\operatorname{D} = \operatorname{signal} + \lambda \operatorname{noise}, \label{eq.generative.model}
\end{equation}
where the signal is structured (usually sparse or low rank) and the noise obeys a simple probabilistic model and it is somehow \emph{incoherent} with the signal structure.
The classic mixtures of Gaussians and the currently popular stochastic block model \cite{abbe2017community} can be seen as examples of generative models of form \eqref{eq.generative.model}. When designing a generative model for a problem of the form \eqref{eq.generic} a natural structure for the signal is one that makes the solution of \eqref{eq.generic} obvious. For instance, the stochastic ball model \cite{nellore2015recovery} was introduced by Nellore and Ward in order to study the performance of a clustering algorithm, and it consists of points drawn from disjoint unit balls, making the solution to the clustering problem apparent.

In the case of the network alignment, we may let $D_{\lambda}= (G_1, G_2:= P^\top G_1 P + E_\lambda )$ where $P$ is an $n\times n$ permutation matrix and $G_1$ may itself obey a generative model such as Erd\H{o}s-Reny\'i. (Here we are abusing the notation and using $G_1$ and $G_2$ the adjacency matrix of said graphs). The matrix $E_\lambda$ is the noise, for instance switching randomly a $\lambda$ fraction of the edges (from 0 to 1 or from 1 to 0). For $\lambda=0$ ($E_\lambda = 0$) a solution of the network alignment problem is the permutation $P$, however the solution is not necessarily unique. In this context the permutation $P$ is considered to be the \emph{planted solution} or \emph{ground truth} for the alignment problem. If the noise level $\lambda$ is small and $G_1$ is well-conditioned (namely $G_1$ is a sufficiently large asymmetric graph) one expects that $P$ continues to be a solution of \eqref{eq.generic} for $D_{\lambda}$ until some threshold $\lambda^\star(G_1)$ where $P$ stops being the solution of \eqref{eq.generic} but maybe the solution $P^*$ is correlated with $P$. When $\lambda$ is too large then we expect $P^*$ not to contain meaningful information about $P$,
and it makes no sense to think of $P$ as the planted solution anymore. These statistical thresholds for recovery and weak recovery are classic statistical questions and had been studied for the stochastic block model \cite{abbe2017community}. For the graph matching problem there exists partial results \cite{onaran2016optimal} but the problem is still open.

A completely different question is whether one can find the solution of \eqref{eq.generic} in polynomial-time. An active research area in Theoretical Computer Science and Statistics is to understand the interplay between
statistical and computational detection thresholds for community detection problem\cite{abbe2017community}. See \cite{berthet2013complexity, chandrasekaran2013computational} for other examples of statistical inference tradeoffs under computational constraints.
Algorithms studied in this context include spectral methods, projected power iteration and approximate message passing. In some restrictive cases, these algorithms are proven to achieve recovery all the way to the threshold where recovery is possible. There is a very nice explanation of this phenomenon and how different algorithms relate to each other by Perry and collaborators\cite{perry2016message}. Another line of algorithms that has been explored towards these problems are automatically learned algorithms using data-driven techniques like neural networks\cite{bruna2017community, nowak2017note}. One issue to stress is that algorithms designed to address problem \eqref{eq.generic} under a certain model of the form \eqref{eq.generative.model} may be over fitting the statistical model and may not work under small modifications as it has recently been observed using semi-random models\cite{moitra2016robust}.

Returning to the network alignment problem and the model $D_\lambda$ we defined, at $\lambda=0$ the question is equivalent to the graph isomorphism problem for which Babai recently has announced a quasi-polynomial algorithm\cite{babai2016graph}. However, depending on the generative model, the problem can be much simpler like for friendly graphs \cite{aflalo2015convex}. Efficient heuristic algorithms of graph matching for special applications are available without theoretical guarantees and the reader is referred to comprehensive surveys\cite{conte2004thirty}. In this paper we define Projected Power Alignment, a projected power iteration algorithm based on EigenAlign. EigenAling is a spectral method recently introduced by Feizi and collaborators\cite{feizi2016spectral} based on an  relaxation of the quadratic assignment problem.

 \subsection*{Our contribution}
In this paper we propose an algorithm, Projected Power Alignment, which is a projected power iteration version for the spectral algorithm EigenAlign\cite{feizi2016spectral}. We numerically compare its performance with EigenAlign for the Erd\H{o}s-Reny\'i with noise generative model. 

Future work includes the study of the evolution equation for Projected Power Alignment, study of the statistical thresholds for recovery for the Erd\H{o}s-Reny\'i model and an approximate message passing version of our algorithm. 

\subsection*{Organization of this paper}
In Section \ref{sec:qap} we review the quadratic assignment problem, and in Section \ref{sec:eigen} we explain the algorithm EigenAlign\cite{feizi2016spectral} which we use as a baseline for our projected power iteration alignment algorithm. In Section \ref{sec:ppi} we describe projected power iteration algorithms. In Section \ref{sec:algorithm} we introduce our projected power iteration for alignment. In Section \ref{sec:numerics} we present numerical simulations.

\subsection{Graph matching and Quadratic assignment}
\label{sec:qap}

Graph matching is a classic problem in theoretical computer science. Given two graphs with $n\times n$ adjacency matrices $G_1,G_2$, the graph matching problem asks for the best permutation matrix to \emph{align} them. Mathematically it can be expressed as \begin{equation} \min_{X\in \Pi} \|G_1X-XG_2\|_F^2,
\label{eq.matching}
\end{equation}
where $\Pi$ is the set of all permutation matrices of size $n\times n$. Graph matching is closely related with the quadratic assignment problem, that has the graph isomorphism, traveling salesman problem, Gromov-Hausdorff distance\cite{villar2016polynomial} and the minimum bisection as particular cases.

The quadratic assignment problem (QAP) and graph isomorphism were deeply investigated in computer science literature from a worst-case point of view. QAPs in general are known to be NP-hard to solve exactly and even to approximate, no polynomial time algorithm can be guaranteed to approximate the solution up to a small constant factor \cite{queyranne1986performance, makarychev2010maximum}. Graph isomorphism problem is a very special case of QAP, that in our case asks the existence of such $X\in \Pi$ that vanishes \eqref{eq.matching}. It is not known whether the graph isomorphism problem is in P or not, despite the recent advancement that proposed a quasi-polynomial time algorithm \cite{babai2016graph}.

The data science approach has shifted the focus from worst-case performance to average-case complexity. Namely finding a polynomial time algorithm that works for most cases. For instance, community detection, although classically studied as NP-hard min-bisection problem, recently revealed interesting information-theoretic analysis and efficient algorithms with high probability guarantees under \emph{stochastic block model} \cite{abbeexact, massoulie2014community, mossel2013proof}.   

Our problem in the random case can be described as follows. Assume we have two unlabeled graphs $G_1$ and $G_2$, that are noisy observations from an underlying graph. Noise here refers to edge addition or deletion. Since $G_1$ and $G_2$ are observations of the same graph they induce a planted bijection across the node sets of the two graphs and our goal is to recover it. Equivalently it can be stated as having observed two adjacency graphs $G_1$ and $G_2$ where $G_2=PG_1P^T+W$ for some $P\in \Pi$ and noise matrix $W$, can we recover $P$? Note that the maximum a posteriori estimate of $P$ is the solution of graph matching problem \eqref{eq.matching} provided independent small noise. 

Let us consider $G_1$ to be an from Erd\H{o}s-R\'enyi graph on $n$ nodes and edges independently drawn with probability $p$. In the noiseless case (the isomorphism problem on random graphs from ER($n,p$) distribution) the problem can be efficiently solved with high probability \cite{babai1980random, czajka2008improved} by canonical labeling provided  $p\in [\Theta(\frac{\ln n}{n}),1-\Theta(\frac{\ln n}{n})]$. This range of $p$ covers the region of almost sure asymmetry for ER($n,p$) by a constant factor \cite{erdHos1963asymmetric}. Therefore there are efficient algorithms that are order optimal.
Despite the success with high probability of polynomial-time algorithms in graph isomorphism, there is no known efficient algorithm for the general case ($W\neq 0$) that is of satisfying success with respect to information theoretic bounds. In fact, the information theoretic bounds have been only recently studied \cite{onaran2016optimal, cullina2016improved}.  
 
\subsection{The spectral method EigenAlign}
\label{sec:eigen}
Given two unweighted undirected graphs $G_1=(V_1, E_1)$ and $G_2=(V_2,E_2)$ with $n$ vertices each, the EigenAlign algorithm\cite{feizi2016spectral} first defines the alignment matrix $A(G_1,G_2)$. The alignment matrix that is a $n^2\times n^2$ that encodes pairwise information of edges of $G_1$ and $G_2$ giving a score to whether edges match or not. Namely,
\begin{equation} \label{A}
A[(i,j'),(r,s')]=\left\{
\begin{matrix}
s_1 & \text{if }  (i,r)\in E_1  \text{ and } (j',s')\in E_2 & \text{(edge match)}\\
s_2 & \text{if }  (i,r)\not \in E_1  \text{ and } (j',s')\not \in E_2 \\
s_3 & \text{ otherwise} & \text{(edge mismatch).}
\end{matrix}
\right.
\end{equation}
This definition can be extended for weighted or undirected graphs. The network alignment on the matrix $A$ is
\begin{align}
\operatorname{minimize}_{\mathbf{y}}\;& \mathbf{y}^T A \mathbf{y} \label{eq.integer}\\
\nonumber\text{subject to}\;& \sum_{i}y_{i,j'} =1 \quad \text{for all } j'\in V_2 \\
\nonumber & \sum_{j'}y_{i,j'} =1 \quad \text{for all } i\in V_1\\
&\nonumber y_{i,j'}\in \{0,1\} \quad \text{for all } (i,j')\in V_1\times V_2.
\end{align}
Note that the constraints in \eqref{eq.integer} are equivalent to asking $\mathbf{y}$ to be the vectorization of a permutation matrix. Using properties of the Kronecker product we have $\operatorname{Trace}(G_1XG_2X^\top)=\operatorname{vec}(X)^\top G_1\otimes G_2 \operatorname{vec}(X)$ and one can check that 
$$\mathbf{y}^\top A \mathbf{y} = \operatorname{Trace}(G_1'XG_2'X^\top) + \text{constant}$$
where $\operatorname{vec}(X)=\mathbf{y}$ and 
\begin{align}
G_1'&= G_1 + \frac{s_3-s_2}{s1+s_2-2s_3}\\
G_2'&= (s_1+s_2-2s_3)G_2 +(s_2-s_3)
\end{align}
The EigenAlign algorithm\cite{feizi2016spectral} reads as follows:

\fbox{\parbox{0.9\textwidth}{
\begin{enumerate}
\item[0.] Compute $A$ as in \eqref{A}.
\item Compute $\mathbf{v}$ the top eigenvector of $A$.
\item Solve the maximum weight bipartite graph matching by solving the following optimization:
\begin{align}
\operatorname{maximize}_{\mathbf{y}} \;& \mathbf{v}^\top\mathbf{y}, \label{eq.bipartite} \\
\text{subject to}\;& \sum_{i} y_{i,j'}=1 \quad \text{for all } j'\in V_2 \nonumber \\
&\sum_{j'} y_{i,j'}=1 \quad \text{for all } i\in V_1 \nonumber \\
&0\leq y_{i,j'}\leq 1 \quad \text{for all } (i,j')\in V_1\times V_2 \nonumber
\end{align}
\end{enumerate}
}
}

Note the solution of \eqref{eq.bipartite} is always integral (i.e. $y_{i,j'}\in\{0,1\}$ for all $(i,j')\in V_1\times V_2$) because \eqref{eq.bipartite} is a totally unimodular linear program\cite{garfinkel1972integer}. The parameters suggested for EigenAlign are $s_1=\alpha+\epsilon$, $s_2=1+\epsilon$, $s_3=\epsilon$, for $\epsilon=0.001$ and $$\alpha= 1 + \frac{\#\text{matches}}{\#\text{mismatches}}.$$ 
This choice of parameters is motivated by keeping the weights of matches and mismatches balanced in the objective function and observed to lead to better performance as reported in Feizi's paper.

\subsection{Projected power iteration}
\label{sec:ppi}

Projected power iteration method was proposed by Journ\'ee et.al. for sparse PCA problem\cite{journee2010generalized} and at each iteration $t$, it can simply be expressed as
\begin{align}
    \mathbf v_{t+1} &= \mathcal F (\mathbf{Av}_t)
\end{align}
where $\mathbf A$ is the matrix whose leading eigenvector is to be computed and $\mathcal F$ is a non-linear projection. The starting point $v_0$ can be chosen arbitrarily.

At the same time as projected power method was introduced another branch of algorithms with a very similar idea, called iterative thresholding, was popular in compressed sensing applications \cite{daubechies2004iterative}. Although much more efficient than convex $\ell_1$ relaxation algorithms, iterative thresholding was not as accurate. The improvement on iterative thresholding that equalized their accuracy to convex relaxation without sacrificing efficiency is achieved by adding a correction term to each iterate, called \emph{onsager term} in statistical physics. The motivation for the onsager term is due to belief propagation algorithms\cite{donoho2009message} and hence the algorithm with corrected iterates is called \emph{approximate message passing} (AMP). Projected power method or AMP have been successfully applied to eigenproblems in different contexts, each of them taking advantage of the specific structure of the set in which the problem is defined in the projection step of the algorithm. These include constrained PCA \cite{deshpande2014cone}, \cite{montanari2016non}, phase synchronization \cite{boumal2016nonconvex, perry2016message} and community detection \cite{deshpande2015asymptotic, chen2016projected}. For some generative random models AMP is shown to achieve statistical upper bounds on accuracy by inspection of \emph{state evolution} equations\cite{montanari2016non}.

\section{Projected Power Alignment}
\label{sec:algorithm}

The algorithm we propose, Projected Power Alignment, constructs the alignment matrix $A$ in the same fashion as EigenAlign \eqref{A}, and likewise EigenAlign computes the top eigenvector of $A$ as an initialization step. Instead of rounding the top eigenvector to a permutation matrix with a linear program, Projected Power Alignment does a \emph{projection step} by rounding the vector into a permutation in a greedy fashion (for efficiency purposes) and then iteratively alternates between multiplying by $A$ and rounding. 

\fbox{
\parbox{0.9\textwidth}{
\begin{enumerate}
\item[0.] Compute $A$ as in \eqref{A}.
\item Let $\mathbf{v}^0$ the top eigenvector of $A$.
\item For $t=1,\ldots$ until convergence or time out
\begin{enumerate}
\item Let $\mathbf u^{t+1}=A\mathbf v^t$.
\item Greedily round $\mathbf{u}^{t+1}$ into a permutation matrix by rounding the largest entry of $\mathbf u_{i,j'}^{t+1}$ to 1 and setting the remaining elements of its row and column to 0 iteratively until we obtain the permutation matrix $\mathbf v^{t+1}=\mathcal F(\mathbf u^{t+1})$.
\end{enumerate}
\end{enumerate}
}
}

State evolution analysis, as mentioned before, is the key to characterize the performance of power iteration method or AMP. Its hypothesis is that at each iteration $t$ the iterand can be approximated as the optimal solution and an added independent gaussian noise with diminishing variance if SNR of the original problem is large enough. 
This remarkable observation was rigorously proven in \cite{bayati2011dynamics} and \cite{javanmard2013state} in the case the matrix $A$ is assumed to be drawn from a distribution $\mathbf A$ satisfying certain conditions.
Although our problem has important differences than PCA or sparse recovery context, which these proofs are built in, a recent paper shows statistical equivalence of community detection and PCA with gaussian noise, namely guessing the leading eigenvector of $A$ where
\[A= \sqrt{\frac{\lambda}{n}}\mathbf x \mathbf x^T +Z  \]
where $\mathbf x$ is the vector of community assignments (-1 or 1), $ Z$ is gaussian symmetric matrix independent of $\mathbf x$ and $\lambda$ is a function of the difference between intra and inter-community edge probabilities. We suspect such equivalence holds for our problem too so we could use state evolution analysis for AMP of PCA. However we defer this to future work.

\section{Numerical simulations}
\label{sec:numerics}

\begin{figure}
    \centering
    \begin{tabular}{cc}
    \includegraphics[width=.45\textwidth]{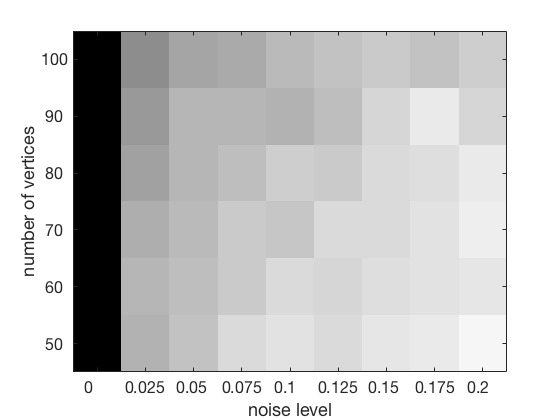} &
    \includegraphics[width=.45\textwidth]{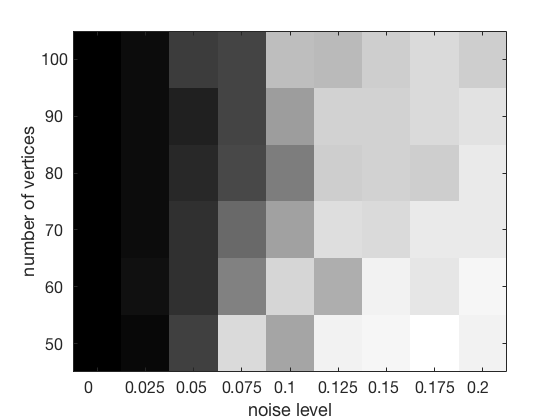} \\
    \includegraphics[width=.45\textwidth]{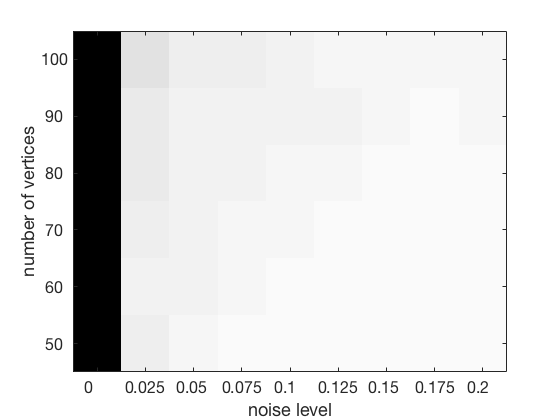} &
    \includegraphics[width=.45\textwidth]{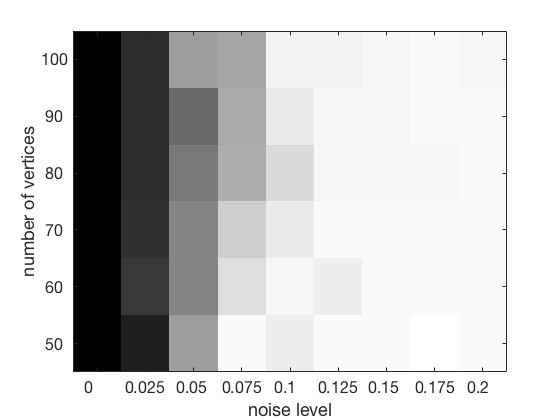} \\
    \includegraphics[width=.45\textwidth]{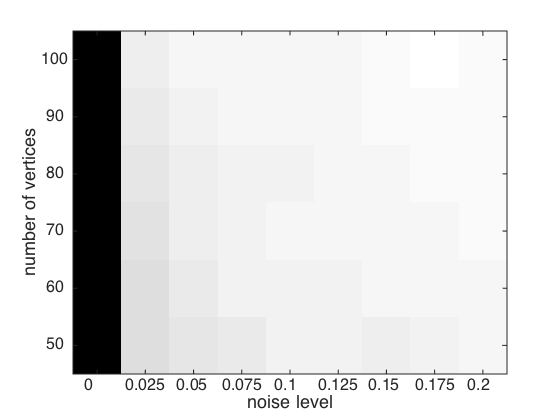} &
    \includegraphics[width=.45\textwidth]{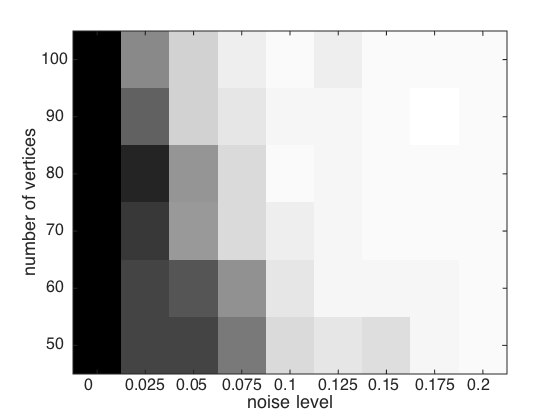}
    \end{tabular}
    \caption{Recovery rates for EigenAlign (left) and Projected Power Alignment (right) for the Erd\H{o}s-Reny\'i with noise model described in Section \ref{sec:numerics}. We take $p=0.2$ (top line) and a logarithm scale for enhacing visualization. We take the same plot of $p=0.2$ with a linear scale (middle line). We take $p=0.5$ and a linear scale (bottom line). Darker color correspond to higher recovery rate. The recovery rate is computed as the average performance among 20 independent experiments. Note that both algorithms obtain perfect recovery rate in the noiseless case. Note that, consistently with the theory by Feizi and collaborators\cite{feizi2016spectral}, the recovery rate seems to (modestly) increase with $n$ for $p=0.2$ but decrease with $n$ for $p=0.5$.}
    \label{fig:phase_fig}
\end{figure}

In Figure \ref{fig:phase_fig} we report on numerical experiments where we compare the recovery rates of Projected Power Alignment with EigenAlign\cite{feizi2016spectral} under the Erd\H{o}s-Reny\'i model. We draw $G_1$ to be an Erd\H{o}s-Reny\'i graph with edge probability $p$ and and $G_2=P^\top \tilde G_1 ( \odot E_{\lambda}) P$ where $\tilde G_1$ is a noisy version of $G_1$ according to the noise model used in the EigenAlign paper \cite{feizi2016spectral}. Namely 
$$\tilde G_1= G_1 \odot (1-Q) + (1-G_1)\odot Q$$
where $\odot$ denotes the Hadamard product, and $Q$ is a symmetric binary random matrix whose edges are
drawn i.i.d. from a Bernoulli distribution with $\mathbb P [Q(i, j) = 1] = \lambda$. We say $\lambda$ is the noise level. 

\acknowledgments
The authors thank Afonso Bandeira for discussions about the problem and Dustin Mixon for reading this draft and providing useful comments. SV is partly supported by the Simons Foundation Algorithms and Geometry (A\&G) Think Tank.
 
\bibliography{report} 

\begin{thebibliography}{10}

\bibitem{yang2007path}
Yang, Q. and Sze, S.-H., ``Path matching and graph matching in biological
  networks,'' {\em Journal of Computational Biology}~{\bf 14}(1),  56--67
  (2007).

\bibitem{rocha1994shape}
Rocha, J. and Pavlidis, T., ``A shape analysis model with applications to a
  character recognition system,'' {\em IEEE Transactions on Pattern Analysis
  and Machine Intelligence}~{\bf 16}(4),  393--404 (1994).

\bibitem{ehrig1992introduction}
Ehrig, H., Habel, A., and Kreowski, H.-J., ``Introduction to graph grammars
  with applications to semantic networks,'' {\em Computers \& Mathematics with
  Applications}~{\bf 23}(6-9),  557--572 (1992).

\bibitem{Kelley30092003}
Kelley, B.~P., Sharan, R., Karp, R.~M., Sittler, T., Root, D.~E., Stockwell,
  B.~R., and Ideker, T., ``Conserved pathways within bacteria and yeast as
  revealed by global protein network alignment,'' {\em Proceedings of the
  National Academy of Sciences}~{\bf 100}(20),  11394--11399 (2003).

\bibitem{goemans1995improved}
Goemans, M.~X. and Williamson, D.~P., ``Improved approximation algorithms for
  maximum cut and satisfiability problems using semidefinite programming,''
  {\em Journal of the ACM (JACM)}~{\bf 42}(6),  1115--1145 (1995).

\bibitem{abbe2017community}
Abbe, E., ``Community detection and stochastic block models: recent
  developments,'' {\em arXiv preprint arXiv:1703.10146}  (2017).

\bibitem{nellore2015recovery}
Nellore, A. and Ward, R., ``Recovery guarantees for exemplar-based
  clustering,'' {\em Information and Computation}~{\bf 245},  165--180 (2015).

\bibitem{onaran2016optimal}
Onaran, E., Garg, S., and Erkip, E., ``Optimal de-anonymization in random
  graphs with community structure,'' in [{\em Signals, Systems and Computers,
  2016 50th Asilomar Conference on}{\nolinebreak\hspace{0.1em}]},   709--713,
  IEEE (2016).

\bibitem{berthet2013complexity}
Berthet, Q. and Rigollet, P., ``Complexity theoretic lower bounds for sparse
  principal component detection,'' in [{\em Conference on Learning
  Theory}{\nolinebreak\hspace{0.1em}]},   1046--1066 (2013).

\bibitem{chandrasekaran2013computational}
Chandrasekaran, V. and Jordan, M.~I., ``Computational and statistical tradeoffs
  via convex relaxation,'' {\em Proceedings of the National Academy of
  Sciences}~{\bf 110}(13),  E1181--E1190 (2013).

\bibitem{perry2016message}
Perry, A., Wein, A.~S., Bandeira, A.~S., and Moitra, A., ``Message-passing
  algorithms for synchronization problems over compact groups,'' {\em arXiv
  preprint arXiv:1610.04583}  (2016).

\bibitem{bruna2017community}
Bruna, J. and Li, X., ``Community detection with graph neural networks,'' {\em
  arXiv preprint arXiv:1705.08415}  (2017).

\bibitem{nowak2017note}
Nowak, A., Villar, S., Bandeira, A.~S., and Bruna, J., ``A note on learning
  algorithms for quadratic assignment with graph neural networks,'' {\em arXiv
  preprint arXiv:1706.07450}  (2017).

\bibitem{moitra2016robust}
Moitra, A., Perry, W., and Wein, A.~S., ``How robust are reconstruction
  thresholds for community detection?,'' in [{\em Proceedings of the
  forty-eighth annual ACM symposium on Theory of
  Computing}{\nolinebreak\hspace{0.1em}]},   828--841, ACM (2016).

\bibitem{babai2016graph}
Babai, L., ``Graph isomorphism in quasipolynomial time,'' in [{\em Proceedings
  of the 48th Annual ACM SIGACT Symposium on Theory of
  Computing}{\nolinebreak\hspace{0.1em}]},   684--697, ACM (2016).

\bibitem{aflalo2015convex}
Aflalo, Y., Bronstein, A., and Kimmel, R., ``On convex relaxation of graph
  isomorphism,'' {\em Proceedings of the National Academy of Sciences}~{\bf
  112}(10),  2942--2947 (2015).

\bibitem{conte2004thirty}
Conte, D., Foggia, P., Sansone, C., and Vento, M., ``Thirty years of graph
  matching in pattern recognition,'' {\em International journal of pattern
  recognition and artificial intelligence}~{\bf 18}(03),  265--298 (2004).

\bibitem{feizi2016spectral}
Feizi, S., Quon, G., Recamonde-Mendoza, M., M{\'e}dard, M., Kellis, M., and
  Jadbabaie, A., ``Spectral alignment of networks,'' {\em arXiv preprint
  arXiv:1602.04181}  (2016).

\bibitem{villar2016polynomial}
Villar, S., Bandeira, A.~S., Blumberg, A.~J., and Ward, R., ``A polynomial-time
  relaxation of the gromov-hausdorff distance,'' {\em arXiv preprint
  arXiv:1610.05214}  (2016).

\bibitem{queyranne1986performance}
Queyranne, M., ``Performance ratio of polynomial heuristics for triangle
  inequality quadratic assignment problems,'' {\em Operations Research
  Letters}~{\bf 4}(5),  231--234 (1986).

\bibitem{makarychev2010maximum}
Makarychev, K., Manokaran, R., and Sviridenko, M., ``Maximum quadratic
  assignment problem: Reduction from maximum label cover and lp-based
  approximation algorithm,'' {\em Automata, Languages and Programming} ,
  594--604 (2010).

\bibitem{abbeexact}
Abbe, E., Bandeira, A.~S., and Hall, G., ``Exact recovery in the stochastic
  block model,'' {\em IEEE Transactions on Information Theory}~{\bf 62},
  471--487 (Jan 2016).

\bibitem{massoulie2014community}
Massouli{\'e}, L., ``Community detection thresholds and the weak ramanujan
  property,'' in [{\em Proceedings of the forty-sixth annual ACM symposium on
  Theory of computing}{\nolinebreak\hspace{0.1em}]},   694--703, ACM (2014).

\bibitem{mossel2013proof}
Mossel, E., Neeman, J., and Sly, A., ``A proof of the block model threshold
  conjecture,'' {\em arXiv preprint arXiv:1311.4115}  (2013).

\bibitem{babai1980random}
Babai, L., Erdos, P., and Selkow, S.~M., ``Random graph isomorphism,'' {\em
  SIaM Journal on computing}~{\bf 9}(3),  628--635 (1980).

\bibitem{czajka2008improved}
Czajka, T. and Pandurangan, G., ``Improved random graph isomorphism,'' {\em
  Journal of Discrete Algorithms}~{\bf 6}(1),  85--92 (2008).

\bibitem{erdHos1963asymmetric}
Erd{\H{o}}s, P. and R{\'e}nyi, A., ``Asymmetric graphs,'' {\em Acta Mathematica
  Hungarica}~{\bf 14}(3-4),  295--315 (1963).

\bibitem{cullina2016improved}
Cullina, D. and Kiyavash, N., ``Improved achievability and converse bounds for
  erdos-r{\'e}nyi graph matching,'' in [{\em Proceedings of the 2016 ACM
  SIGMETRICS International Conference on Measurement and Modeling of Computer
  Science}{\nolinebreak\hspace{0.1em}]},   63--72, ACM (2016).

\bibitem{garfinkel1972integer}
Garfinkel, R.~S., Nemhauser, G.~L., et~al.,  [{\em Integer
  programming}{\nolinebreak\hspace{0.1em}]}, vol.~4, Wiley New York (1972).

\bibitem{journee2010generalized}
Journ{\'e}e, M., Nesterov, Y., Richt{\'a}rik, P., and Sepulchre, R.,
  ``Generalized power method for sparse principal component analysis,'' {\em
  Journal of Machine Learning Research}~{\bf 11}(Feb),  517--553 (2010).

\bibitem{daubechies2004iterative}
Daubechies, I., Defrise, M., and De~Mol, C., ``An iterative thresholding
  algorithm for linear inverse problems with a sparsity constraint,'' {\em
  Communications on pure and applied mathematics}~{\bf 57}(11),  1413--1457
  (2004).

\bibitem{donoho2009message}
Donoho, D.~L., Maleki, A., and Montanari, A., ``Message-passing algorithms for
  compressed sensing,'' {\em Proceedings of the National Academy of
  Sciences}~{\bf 106}(45),  18914--18919 (2009).

\bibitem{deshpande2014cone}
Deshpande, Y., Montanari, A., and Richard, E., ``Cone-constrained principal
  component analysis,'' in [{\em Advances in Neural Information Processing
  Systems}{\nolinebreak\hspace{0.1em}]},   2717--2725 (2014).

\bibitem{montanari2016non}
Montanari, A. and Richard, E., ``Non-negative principal component analysis:
  Message passing algorithms and sharp asymptotics,'' {\em IEEE Transactions on
  Information Theory}~{\bf 62}(3),  1458--1484 (2016).

\bibitem{boumal2016nonconvex}
Boumal, N., ``Nonconvex phase synchronization,'' {\em SIAM Journal on
  Optimization}~{\bf 26}(4),  2355--2377 (2016).

\bibitem{deshpande2015asymptotic}
Deshpande, Y., Abbe, E., and Montanari, A., ``Asymptotic mutual information for
  the two-groups stochastic block model,'' {\em arXiv preprint
  arXiv:1507.08685}  (2015).

\bibitem{chen2016projected}
Chen, Y. and Candes, E., ``The projected power method: An efficient algorithm
  for joint alignment from pairwise differences,'' {\em arXiv preprint
  arXiv:1609.05820}  (2016).

\bibitem{bayati2011dynamics}
Bayati, M. and Montanari, A., ``The dynamics of message passing on dense
  graphs, with applications to compressed sensing,'' {\em IEEE Transactions on
  Information Theory}~{\bf 57}(2),  764--785 (2011).

\bibitem{javanmard2013state}
Javanmard, A. and Montanari, A., ``State evolution for general approximate
  message passing algorithms, with applications to spatial coupling,'' {\em
  Information and Inference: A Journal of the IMA}~{\bf 2}(2),  115--144
  (2013).

\end{thebibliography}
\bibliographystyle{spiebib} 

\end{document}